\documentclass[12pt]{article}
\usepackage{fancyhdr}

\usepackage{mathrsfs}
\usepackage[T1]{fontenc}
\usepackage{mathpazo}
\usepackage{setspace}
\usepackage{amsfonts}
\usepackage{amssymb}
\usepackage{amsmath}
\usepackage{epsfig}
\usepackage{latexsym}
\usepackage{color}
\usepackage{graphicx}
\usepackage{nicefrac}
\usepackage[latin1]{inputenc}
\usepackage{slashed}
\usepackage{multirow}
\usepackage{comment}
\usepackage{soul}
\usepackage{hyperref}
\usepackage[nosort]{cite}
\usepackage{datetime}
\usepackage{braket}

\usepackage[titletoc]{appendix}

\def\hybrid{\topmargin -20pt    \oddsidemargin 0pt
        \headheight 0pt \headsep 0pt
        \textwidth 6.25in       
        \textheight 9.25in       
        \marginparwidth .875in
        \parskip 5pt plus 1pt   \jot = 1.5ex}

\hybrid

\def\baselinestretch{1.2}

\catcode`\@=11

\def\marginnote#1{}
\newcount\hour
\newcount\minute
\newtoks\amorpm
\hour=\time\divide\hour by60
\minute=\time{\multiply\hour by60 \global\advance\minute by-\hour}
\edef\standardtime{{\ifnum\hour<12 \global\amorpm={am}%
        \else\global\amorpm={pm}\advance\hour by-12 \fi
        \ifnum\hour=0 \hour=12 \fi
        \number\hour:\ifnum\minute<10 0\fi\number\minute\the\amorpm}}
\edef\militarytime{\number\hour:\ifnum\minute<10 0\fi\number\minute}

\def\draftlabel#1{{\@bsphack\if@filesw {\let\thepage\relax
   \xdef\@gtempa{\write\@auxout{\string
      \newlabel{#1}{{\@currentlabel}{\thepage}}}}}\@gtempa
   \if@nobreak \ifvmode\nobreak\fi\fi\fi\@esphack}
        \gdef\@eqnlabel{#1}}
\def\@eqnlabel{}
\def\@vacuum{}
\def\draftmarginnote#1{\marginpar{\raggedright\scriptsize\tt#1}}

\def\draft{\oddsidemargin -.5truein
        \def\@oddfoot{\sl preliminary draft \hfil
        \rm\thepage\hfil\sl\today\quad\militarytime}
        \let\@evenfoot\@oddfoot \overfullrule 3pt
        \let\label=\draftlabel
        \let\marginnote=\draftmarginnote
   \def\@eqnnum{(\theequation)\rlap{\kern\marginparsep\tt\@eqnlabel}%
\global\let\@eqnlabel\@vacuum}  }

\def\preprint{\twocolumn\sloppy\flushbottom\parindent 2em
        \leftmargini 2em\leftmarginv .5em\leftmarginvi .5em
        \oddsidemargin -.5in    \evensidemargin -.5in
        \columnsep .4in \footheight 0pt
        \textwidth 10.in        \topmargin  -.4in
        \headheight 12pt \topskip .4in
        \textheight 6.9in \footskip 0pt
        \def\@oddhead{\thepage\hfil\addtocounter{page}{1}\thepage}
        \let\@evenhead\@oddhead \def\@oddfoot{} \def\@evenfoot{} }

\def\numberbysection{\@addtoreset{equation}{section}
        \def\theequation{\thesection.\arabic{equation}}}

\def\underline#1{\relax\ifmmode\@@underline#1\else
        $\@@underline{\hbox{#1}}$\relax\fi}

\def\titlepage{\@restonecolfalse\if@twocolumn\@restonecoltrue\onecolumn
     \else \newpage \fi \thispagestyle{empty}\c@page\z@
        \def\thefootnote{\fnsymbol{footnote}} }

\def\endtitlepage{\if@restonecol\twocolumn \else \newpage \fi
        \def\thefootnote{\arabic{footnote}}
        \setcounter{footnote}{0}}  

\catcode`@=12
\relax

\def\figcap{\section*{Figure Captions\markboth
        {FIGURECAPTIONS}{FIGURECAPTIONS}}\list
        {Figure \arabic{enumi}:\hfill}{\settowidth\labelwidth{Figure
999:}
        \leftmargin\labelwidth
        \advance\leftmargin\labelsep\usecounter{enumi}}}
 \relax
\def\tablecap{\section*{Table Captions\markboth
        {TABLECAPTIONS}{TABLECAPTIONS}}\list
        {Table \arabic{enumi}:\hfill}{\settowidth\labelwidth{Table
999:}
        \leftmargin\labelwidth
        \advance\leftmargin\labelsep\usecounter{enumi}}}
 \relax
\def\reflist{\section*{References\markboth
        {REFLIST}{REFLIST}}\list
        {[\arabic{enumi}]\hfill}{\settowidth\labelwidth{[999]}
        \leftmargin\labelwidth
        \advance\leftmargin\labelsep\usecounter{enumi}}}
 \relax
\makeatletter
\newcounter{pubctr}
\def\publist{\@ifnextchar[{\@publist}{\@@publist}}
\def\@publist[#1]{\list
        {[\arabic{pubctr}]\hfill}{\settowidth\labelwidth{[999]}
        \leftmargin\labelwidth
        \advance\leftmargin\labelsep
        \@nmbrlisttrue\def\@listctr{pubctr}
        \setcounter{pubctr}{#1}\addtocounter{pubctr}{-1}}}
\def\@@publist{\list
        {[\arabic{pubctr}]\hfill}{\settowidth\labelwidth{[999]}
        \leftmargin\labelwidth
        \advance\leftmargin\labelsep
        \@nmbrlisttrue\def\@listctr{pubctr}}}
 \relax
\makeatother
\newskip\humongous \humongous=0pt plus 1000pt minus 1000pt

\newif\ifdtup

\relax

\def\be{\begin{equation}}
\def\ee{\end{equation}}
\def\ba{\begin{eqnarray}}
\def\ea{\end{eqnarray}}

\def\del{\partial}

\def\k{\kappa}

\def\b{\beta}

\def\g{\gamma}

\def\d{\delta}
\def\D{\Delta}

\def\om{\omega}
\def\Om{\Omega}
\def\l{\lambda}
\def\L{\Lambda}
\def\s{\sigma}

\def\no{\noindent}

\def\qq{\qquad}

\def\IR{\relax{\rm I\kern-.18em R}}

\def \ov {\over}

\def\diag{{\rm diag}}

\def\IR{\relax{\rm I\kern-.18em R}}
\def\IL{\relax{\rm I\kern-.18em L}}

\def\inv{^{\raise.15ex\hbox{${\scriptscriptstyle -}$}\kern-.05em 1}}

\def\Tr{{\rm Tr}}

\begin{document}

\renewcommand{\theequation}{\thesection.\arabic{equation}}
\csname @addtoreset\endcsname{equation}{section}

\newcommand{\beq}{\begin{equation}}
\newcommand{\eeq}[1]{\label{#1}\end{equation}}
\newcommand{\ber}{\begin{equation}}
\newcommand{\eer}[1]{\label{#1}\end{equation}}
\newcommand{\eqn}[1]{(\ref{#1})}
\begin{titlepage}
\begin{center}

\hfill  CERN-TH-2019-053
\vskip .4 cm

\vskip .3 in

{\large\bf An exact symmetry in $\lambda$-deformed CFTs}

\vskip 0.4in
{\bf George Georgiou},$^a$\ {\bf Eftychia Sagkrioti},$^a$\\
\vskip .09 cm
{\bf Konstantinos Sfetsos}$^a$\ and {\bf Konstantinos Siampos}$^{a,b}$
\vskip 0.17in

{\em${}^a$Department of Nuclear and Particle Physics,\\ Faculty of Physics, National and Kapodistrian University of Athens,\\15784 Athens, Greece
}

\vskip 0.1in

{\em${}^b$Theoretical Physics Department, CERN, 1211 Geneva 23, Switzerland}

\vskip 0.1in

{\footnotesize \texttt george.georgiou, esagkrioti, ksfetsos, konstantinos.siampos@phys.uoa.gr}


\vskip .5in
\end{center}

\centerline{\bf Abstract}

\no
We consider $\lambda$-deformed current algebra CFTs at level $k$, interpolating between an exact CFT in the UV and a PCM in the IR. By employing gravitational techniques, we derive the two-loop, in the large $k$ expansion, $\beta$-function. We find that this is covariant under a remarkable exact symmetry involving the coupling $\lambda$, the level $k$ and the adjoint quadratic Casimir of the group. Using this symmetry and CFT techniques, we are able to compute the Zamolodchikov metric, the anomalous dimension of the bilinear operator and the Zamolodchikov $C$-function at two-loops in the large $k$ expansion, as exact functions of the deformation parameter. Finally, we extend the above results to $\lambda$-deformed parafermionic algebra coset CFTs which interpolate between exact coset CFTs in the UV and a symmetric coset space in the IR.

\noindent

\no

\vskip .4in
\noindent
\end{titlepage}
\vfill
\eject

\newpage

\tableofcontents

\noindent

\def\baselinestretch{1.2}
\baselineskip 20 pt
\noindent


\setcounter{equation}{0}
\renewcommand{\theequation}{\thesection.\arabic{equation}}

\section{Introduction}

The classical actions of field theories may easily have certain global symmetries depending on the field content and on the particular form of the constants coupling the various fields.
Discovering emergent non-perturbative symmetries in quantum field theories acting also
in their coupling space can be of major importance.
These may arise unexpectedly and can provide strict constraints on the observables of the theory. An important example of the above is the maximally supersymmetric field theory, $\mathcal N=4$ SYM, which possesses a remarkable non-perturbative symmetry, similar in a sense to the exact symmetry presented  in this work, called S-duality \cite{S-duality}, i.e.
for zero  theta angle this reads $g_{\rm YM}\to 1/g_{\rm YM}$.

\no
There is a certain class of quantum field theories where one may test these ideas and which in recent years have been intensively explored.
In particular, consider a current algebra theory at level $k$ realized by a two-dimensional $\s$-model action, e.g. a WZW model theory \cite{Witten:1983ar} perturbed by current bilinear terms of the form  $\l_{ab} J_+^a J_-^b$. Here, the $\l_{ab}$'s are couplings and elements of a matrix and $a,b$ run over the dimensionality of the Lie-algebra of a semisimple group $G$.
As it stands the action may have certain global symmetries depending on the particular form of
$\l_{ab}$. However, another symmetry appears at the quantum level.
Specifically, it was argued using path integrable techniques \cite{Kutasov:1989aw}, that
the theory is quantum mechanically invariant under an additional remarkable master symmetry.
In the space of couplings this acts as $\l\to \l^{-1}$ and $k\to- k $, where for the purposes of our  introduction we have presented it for $k\gg 1$. This is a non-perturbative symmetry, not valid at any finite order in perturbation theory in the couplings $\l_{ab}$.

The first class of theories where the above symmetry was explicitly realized classically in
a $\s$-model was constructed in \cite{Sfetsos:2013wia}, whereas the symmetry itself was noticed
and demonstrated in \cite{Itsios:2014lca,Sfetsos:2014jfa}.
This action captures all loop effects
in the deformation matrix $\l$ and is valid to leading order for large $k$.
This effective action, in conjunction with results from conformal perturbation theory and the above symmetry has been instrumental
in extracting vast information at the quantum regime of the theory \cite{Georgiou:2015nka}.
This includes the $\beta$-function and the anomalous dimensions of current, primary \cite{Georgiou:2016iom} and composite operators \cite{Georgiou:2016zyo}.
The prototype $\l$-deformed $\s$-model action of \cite{Sfetsos:2013wia} represents
the exact deformation of a single WZW current algebra theory due to the interactions of currents belonging
to the theory, i.e. self-interactions.
Since then, this construction has been extended to cover cases with more than one
current algebra theories, mutually and/or
self-interacting \cite{Georgiou:2017jfi,Georgiou:2016urf,Georgiou:2018hpd,Georgiou:2018gpe }.\footnote{We
mention in passing that perhaps the major
reason these models have attracted attention is integrability. Such cases exist first for isotropic deformation matrices
\cite{Sfetsos:2013wia,Georgiou:2017jfi,Georgiou:2018hpd,Georgiou:2018gpe}
(for the $SU(2)$ group case, integrability has been demonstrated in \cite{Balog:1993es}). Nevertheless
integrability holds for some anisotropic models as well. In particular, for the $\l$-deformed $SU(2)$ based models in \cite{Sfetsos:2014lla,Sfetsos:2015nya}, as well as for subclasses of those in \cite{Georgiou:2018hpd,Georgiou:2018gpe}. Integrable deformations based on cosets, symmetric and semi-symmetric spaces have also been constructed in \cite{Sfetsos:2013wia,Sfetsos:2017sep}, \cite{Hollowood:2014rla} and \cite{Hollowood:2014qma}, respectively.
Finally, deformed models of low dimensionality were promoted to solutions of type-II supergravity \cite{Sfetsos:2014cea,Demulder:2015lva,Borsato:2016zcf,Chervonyi:2016ajp,Borsato:2016ose}.
}
 Compared to the single $\l$-deformed model these models involve several deformation parameters and their renormalization group has a very rich structure, namely their RG flow possesses several fixed points. 
The use of non-trivial outer automorphisms in this context was put forward in \cite{Driezen:2019ykp} for the case of a single group $G$. Outer automorphisms for the product group $G\times G$ was considered earlier in \cite{Georgiou:2016urf}.
In all cases there is an analog of the above mentioned master symmetry involving the levels of
the current algebra and the various deformation matrices \cite{Kutasov:1989aw}.

\no
The next crucial question is how to proceed deeper into the quantum regime of these theories
by going beyond the leading expressions for large $k$, that is go higher in the $\nicefrac1k$ expansion.
Experience shows that perhaps we may progress in computing by brute force the $\beta$-function to two-loops, but unless we understand the fate of the above master symmetry when
such corrections are taken into account, the progress will stay minimal.
The major purpose of the present paper is to precisely make progress along the above
line of research.

The outline of this paper is as follows: In Sec. \ref{into.lambda.models}, we review the $\l$-deformed models constructed in \cite{Georgiou:2017jfi},
which has two interesting limits -- the PCM and the pseudo-chiral model. In Sec. \ref{singlegeometry} we present the two-loop RG flows in the group case for an isotropic coupling. We present a symmetry of the $\b$-functions in the coupling space $(\l,k)$ (\S~\ref{symmetryisosection}).
Using the symmetry and CFT input we determine the Zamolodchikov metric of the current bilinear driving the conformal perturbation (\S~\ref{group.Zamolo}).
Then we work out the Zamolodchikov $C$-function and the anomalous dimension of the current bilinear (\S~\ref{isoCfunction}).
Using the above we determine the Zamolodchikov $C$-function for the $\l$-deformed $G_k$ (\S~\ref{group.connection}).
 In Sec. \ref{coset}, we generalize the above for the coset space $\displaystyle \frac{SU(2)_k\times SU(2)_k}{U(1)_k}$, working out the two-loop $\b$-function and the corresponding symmetry in the coupling space $(\l,k)$. Using the symmetry and CFT data we determine the Zamolodchikov metric of the parafermionic bilinear driving the conformal perturbation (\S~\ref{coset.Zamo}), the Zamolodchikov $C$-function and the anomalous dimension of the parafermionic
bilinear (\S~\ref{cfunction.double.parafermionic}). Using the above we work out the Zamolodchikov $C$-function for the $\l$-deformed $SU(2)_k/U(1)_k$ (\S~\ref{coset.connection}). In Sec. \ref{conclu}, contains some concluding remarks. In App. \ref{RGequalappend} we compute the
two-loop RG flows for the group case at unequal levels. At equal levels it yields the result analyzed in Sec. \ref{singlegeometry} and agreement with
the corresponding limits already described in Sec. \ref{into.lambda.models} is found for the PCM and pseudo-chiral model (\S~\ref{limits}).

\subsection*{Note added}
\vskip -.5 cm

Extensive parts of this work,
including the $\beta$-function equations for the group and coset cases \eqref{betaonetwo}
and \eqref{fjhskdjklswq} below,
have been presented in talks by one of the authors (K. Siampos), at the
Recent Developments in Strings and Gravity (Corfu, Greece, 10-16 September 2019) \cite{Siampos.Corfu} and at the $10^\text{th}$
Crete regional meeting in String Theory (Kolymbari, Greece, 15-22 September 2019) \cite{Siampos.Crete}.
Towards the completion of the present work, the work of \cite{Hoare:2019mcc} appeared where similar issues
concerning the two-loop $\beta$-function in $\l$-deformed models, are discussed.

\section{The $\l$-deformed models}
\label{into.lambda.models}

Consider the following deformed single-level action \cite{Georgiou:2016urf,Georgiou:2017jfi}
\be
 S= S_{k}(\frak{g}_1) + S_{k}(\frak{g}_2)
+ {k\l_{ab}\ov \pi}\, \int  \text{d}^2\s\,J^a_{1+}\,J^b_{2-}\,.
\label{defactigen}
\ee
We have denoted by  $S_{k}(\frak{g})$  the WZW action at level $k$ \cite{Witten:1983ar}
\be
\label{wzwacc}
S_{k}(\frak{g}) = {k\ov 2\pi} \int \text{d}^2\s\, \text{Tr}(\del_+ \frak{g}^{-1} \del_- \frak{g})
+ S_{{\rm WZ},k}(\frak{g})\ ,\quad
S_{{\rm WZ},k}(\frak{g})=
{k\ov 12\pi} \int_\text{B}  \text{Tr}(\frak{g}^{-1} \text{d}\frak{g})^3\ ,
\ee
where $\frak{g}\in G$, with $G$ being a semi-simple group of dimension $\dim G$.
The $t_a$'s are Hermitian matrices normalized to $\text{Tr}\left(t_at_b\right)=\d_{ab}$, $[t_a,t_b]=i f_{abc}t_c$ with $a=1,\dots,\text{dimG}$, where
the structure constants $f_{abc}$ are taken to be real.
The currents $J^a_{\pm}$ are given by
\be
\label{pfkdlddsks}
J_{+}^a=-i\,\text{Tr}\big(t_a\del_+\frak{g}\frak{g}^{-1}\big)\,,\quad J_{-}^a=-i\,\text{Tr}\big(t_a\frak{g}^{-1}\del_-\frak{g}\big)\  .
\ee
We also define the orthogonal matrix $D_{ab}=\text{Tr}\big(t_a\frak{g}t_b\frak{g}^{-1}\big)$.
All these may appear
with an extra index $1$ or $2$ depending on which group element $\frak{g}_1$ or $\frak{g}_2$ will be used in the particular expressions.

\no
The above model can be obtained as a limit of the doubly $\l$-deformed models constructed in \cite{Georgiou:2016urf} - see also \cite{Georgiou:2017jfi} for the unequal level case- by setting 
one of the deformation parameters to zero. In the same works it was also stressed that the linearized action \eqref{defactigen} is, in fact, the effective action incorporating all loop effects in the deformation parameter $\l_{ab}$, that is it does not receive further $\l$-dependent corrections. This is the first reason for using \eqref{defactigen}, instead of the prototype
$\l$-deformed model. The second reason is that, as was shown in \cite{Georgiou:2017aei} by using CFT arguments, both actions share the same  $\beta$-function for the deformation parameter $\l_{ab}$ to all orders not only in the $\l_{ab}$, but also in the $\nicefrac1k$ expansion. 
This is strictly true only when we choose the chiral anti-chiral current two-point function to vanish. 
It is important to note that this is precisely  the choice in which the symmetry of \cite{Kutasov:1989aw} is realized. 
The third reason is that, the $\s$-model \eqref{defactigen} does not receive quantum corrections 
in contradistinction to the action of the single $\l$-deformed model.

The action \eqref{defactigen} has two interesting limits for $\l_{ab}\to\pm\d_{ab}$. They will give rise to the PCM and pseudo-chiral models, respectively.
To analyze the limit $\l_{ab}\to\delta_{ab}$ we rewrite \eqref{defactigen} as
\be
\label{jfjkfhsjhdfjs}
S=S_k\left(\frak{g}_2\frak{g}_1\right)+\left(\l_{ab}-\d_{ab}\right)\int\text{d}^2\s J_{1+}^a J_{2-}^b\ ,
\ee
where we made use of the Polyakov--Wiegmann (PW) identity \cite{Polyakov:1983tt}.\footnote{In our conventions the PW identity reads
\begin{equation*}
S_{k}(\frak{g}_2\frak{g}_1)=S_{k}(\frak{g}_1)+S_{k}(\frak{g}_2)+\frac{k}{\pi}\int\text{d}^2\s\, J_{1+}^aJ_{2-}^a\,.
\end{equation*}
}
Then we perform the following zoom-in limit
\be
\label{pcmlim}
\l_{ab}=\d_{ab}-\frac{E_{ab}}{k}\,,\quad k\gg1\,,\quad \frak{g}_1=\frak{g}^{-1}_2\left(\mathbb{I}+i\frac{u^at_a}{\sqrt{k}}\right)+\cdots\ .
\ee
Then, the action \eqref{jfjkfhsjhdfjs} takes the form of a PCM model, with the $\dim G$
additional spectators bosons $u^a$
\be
\label{PCMlimit}
S_\text{PCM}=-\frac{E_{ab}}{\pi}\int\text{d}^2\s\, \Tr(t^a \frak{g}^{-1}_2 \del_+ \frak{g}_2)
\Tr(t^b \frak{g}^{-1}_2 \del_- \frak{g}_2) + \frac{1}{2\pi}\int\text{d}^2\s\,\del_+u^a\del_-u^a \,.
\ee
We note here that similar to \eqref{pcmlim} a zoom-in limit to the prototype $\l$-deformed action of
\cite{Sfetsos:2013wia}
gives rise to the non-Abelian T-dual of  the PCM $\s$-model. This fact is not a surprise since \eqref{defactigen} is
canonically equivalent \cite{Georgiou:2017oly} to  the sum of a WZW action and the
$\l$-deformed action of \cite{Sfetsos:2013wia}. The two zoom-in limits simply relate the PCM model and its non-Abelian
T-dual which are also known to be canonically equivalent as well  \cite{Curtright:1994be,Alvarez:1994wj}.
This limit is a way to make sense of the theory in the IR when $\l$ approaches unity and  strong coupling 
effects prevail.

\no
To analyze the limit $\l_{ab}\to-\delta_{ab}$ we rewrite \eqref{defactigen} by making use of the PW identity, as
\be
\label{jfjkfhsjhdfjss}
S=S_k\big(\frak{g}_2\frak{g}_1^{-1}\big)+2S_{\text{WZ},k}\left(\frak{g}_1\right)+
\frac{k}{\pi}\int\text{d}^2\s \left(D_1+\l\right)_{ab}J_{1+}^aJ_{2-}^b\ .
\ee
Next, we perform the following slightly different  zoom-in limit
\be
\begin{split}
\label{psliim}
&\l_{ab}=-\d_{ab}+\frac{E_{ab}}{k^{1/3}}\ ,\quad k\gg1\ ,
\\
&\frak{g}_1=\mathbb{I}+i\frac{v^at_a}{2k^{1/3}}-i\frac{u^at_a}{2k^{1/2}}+\cdots\ ,
\quad \frak{g}_2=\mathbb{I}+i\frac{v^at_a}{2k^{1/3}}+i\frac{u^at_a}{2k^{1/2}}+\cdots\ .
\end{split}
\ee
Then, the action \eqref{jfjkfhsjhdfjss} takes the form of the generalized
pseudo-chiral model found in \cite{Georgiou:2016iom} by performing in the prototype $\l$-deformed action a similar to
\eqref{psliim} zoom-in limit plus the $\dim G$ spectator bosons $u^a$
\be
\label{pseudolimit}
S_\text{pseudo}=\frac{1}{4\pi}\int\text{d}^2\s\, \Big(E_{ab}+\frac13f_{ab}\Big)\del_+v^a\del_-v^b
+ \frac{1}{2\pi}\int\text{d}^2\s\,\del_+u^a\del_-u^a\ ,
\ee
where $f_{ab}=f_{abc}v^c$. For diagonal $E_{ab}$ the first term is the prototype pseudo-dual model studied in
\cite{Nappi:1979ig}.
These limits should be well defined at the level of the physical quantities of the theory,
such as for the $\b$-functions and the operator's anomalous dimensions.

\section{The group space}
\label{singlegeometry}

We would like to compute the RG flow equations of \eqref{defactigen} at two-loop order in the $\nicefrac1k$ expansion for isotropic coupling $\l_{ab}=\l\d_{ab}$.
This is a rather long but quite standard computation
that is performed in the App. \ref{RGequalappend}. The end result is that the model is renormalizable at order $\nicefrac{1}{k^2}$ and that there is
no need for a diffeomorphism or an addition of a counter term.
The $\b$-function for $\l$ reads \eqref{betaonetwoappend}
\be
\label{betaonetwo}
\b^\l(\l)={\text{d}\l\ov \text{d}t}=  -{c_G\ov 2 k} {\l^2\ov (1+\l)^2} +
{c_G^2\ov 2 k^2 } {\l^4(1-2\l)\ov (1-\l)(1+\l)^5}\,,
\ee
where $t=\ln\mu^2$, $\mu$ is the RG scale and $c_G$ is the quadratic Casimir in the adjoint representation of the semi-simple group $G$,
i.e. $f_{acd} f_{bcd}=c_G \d_{ab}$.
The level $k$ does not run, thus retaining its topological nature (also) at two-loop order. The above $\b$-function is
well defined in the two interesting zoom-in limits around $\l=\pm1$ performed in the previous section.
These are studied in \S~\ref{limits}.

\subsection{Symmetry}
\label{symmetryisosection}

It has been conjectured  \cite{Kutasov:1989aw} that beyond the leading in the $\nicefrac1k$-expansion, the theory is
invariant under the symmetry
\begin{equation}
\l\to\l^{-1}\,,\quad k\to-k-c_G\,.
\label{cksks}
\end{equation}
It can be easily checked that \eqref{betaonetwo} is not invariant under this to order $\nicefrac1k^2$.
However, contrary to the one-loop result the two-loop result is scheme dependent.
Furthermore,  as was mentioned in \cite{Kutasov:1989aw},  the symmetry \eqref{cksks} is realized  only
 when we choose the chiral anti-chiral current two-point function to vanish.
The fact that the symmetry \eqn{cksks} is not respected by our two-loop $\b$-function indicates that the scheme used in gravity calculations 
is not compatible with the left-right symmetric scheme of the CFT. However,     
it is possible to redefine the coupling $\l$ in such a way that the resulting $\b$-function respects the aforementioned symmetry \eqn{cksks}. 
Based on the general structure of the one-loop in $\nicefrac1k$ results for the $\beta$-function, as well for the
anomalous dimensions of current operators \cite{Georgiou:2015nka}, we redefine $\l$ as
\be
\l=\tilde\l\bigg(1+\frac{c_G}{k}\frac{P(\tilde\l)}{(1-\tilde\l)(1+\tilde\l)^3}\bigg)\ ,
\ee
where $P(\tilde\l)$ is an analytic function of $\tilde\l$. Subsequently, we demand that the symmetry of the $\b$-function becomes
\be
\label{sfjsldjsssk}
\tilde\l\to\tilde\l^{-1}\,,\quad k\to-k-c_G\,.
\ee
This  enforces $P(\tilde\l)$ to satisfy the first-order differential equation
\be
\tilde\l^3 P'(\tilde\l^{-1})-\tilde \l P'(\tilde\l)
+ {\tilde\l^4(\tilde\l^2-3)\ov 1-\tilde\l^2}P(\tilde\l^{-1})+{1-3\tilde\l^2\ov 1-\tilde \l^2}P(\tilde\l)
 +1-\tilde\l^4=0\, .
\ee
This has as a solution the fourth order polynomial
\be
P(\tilde\l)=(1-\tilde\l^2)\big[(1+d_0)\tilde \l^2 + d_1\tilde\l + d_0\big]\ ,
\ee
where $d_{0,1}$ are two arbitrary constants (one is due to the fact that the differential equation involves
$\tilde \l$ as well $\nicefrac{1}{\tilde \l}$ as arguments in $P(\tilde\l)$). Using the above reparametrization  into \eqref{betaonetwo}, we find that
\be
\label{fdjfhshdjs}
\b^{\tilde\l}(\tilde\l)=-\frac{c_G\tilde\l^2}{2k(1+\tilde\l)^2}-\frac{c_G^2\tilde\l^2\big[d_0(1-\tilde\l^2)^2+\tilde\l^2(\tilde\l^2 +2\tilde \l-2)\big]}{2k^2(1-\tilde\l)(1+\tilde\l)^5}\ .
\ee
Note that the constant $d_1$ does not appear in this expression,
while $d_0$ does so and it remains to be determined. To do so first recall again the scheme dependence of the above result concerning the level $k$. We would like to match this scheme to that corresponding to the conformal perturbation theory.
Using the latter, for small $\tilde \l$ the contribution to the $\beta$-function can only be of  ${\cal O}(\nicefrac{\tilde \l^2}{k})$ and a term of
${\cal O}(\nicefrac{\tilde \l^2}{k^2})$ should be absent.
Alternatively, one may establish
that by the fact  that the anomalous dimension of the composite operators $J^a\bar J^a$ is of order one less than the corresponding order of the $\b$-function
(see \eqref{anomalouscompositegeneral} below).
This anomalous dimension cannot have a term of ${\cal O}(\nicefrac{\tilde \l}{k^2})$ since, a linear in $\tilde \l$ term 	arises from a single
insertion operator giving rise to an integral involving the product of a  three-point function of  holomorphic currents with
a similar one with just anti-holomorphic ones. In our normalizations each one of the two correlators contributes a factor of ${\cal O}(\nicefrac{1}{\sqrt{k}})$.
This computation was performed in \cite{Georgiou:2015nka,Georgiou:2016zyo}.
Therefore, one must require the vanishing of the term of  ${\cal O}(\nicefrac{\tilde \l^2}{k^2})$ in \eqref{fdjfhshdjs}.
This can be achieved, for instance, by choosing $d_0=0$ in which case the contribution
of the second term in \eqref{fdjfhshdjs} becomes of ${\cal O}(\nicefrac{\tilde \l^4}{k^2})$. This choice is problematic since it will give rise to non-analytic terms with branch cuts,
i.e. $\ln\frac{1-\l}{1+\l}$,  in the $C$-function as it
will be discussed in the \S~\ref{isoCfunction}.  Their absence implies that $d_0=-\nicefrac12$ which is the choice
we make.
Then, the $\b$-function \eqref{fdjfhshdjs} of course contains a term of  ${\cal O}(\nicefrac{\tilde\l^2}{k^2})$.
To get rid of it we redefine the perturbative parameter from $\nicefrac1k$ to $\nicefrac{1}{k_G}$, where $k_G$ is $k$ shifted by a constant proportional to $c_G$. It turns out that the correct such redefinition is
\be
\label{kgkg}
 k_G=k+ {c_G\ov 2}\ .
 \ee
 Notably, this is the right combination of $k$ and $c_G$ appearing in the Sugawara construction of the energy--momentum tensor in current algebra CFTs and in the conformal dimension of the corresponding primary fields.
Then \eqref{fdjfhshdjs} simplifies to
\be
\boxed{
\label{fdjfhshdjs1}
\b^{\tilde\l}(\tilde\l)=-\frac{c_G\tilde\l^2}{2k_G(1+\tilde\l)^2}-\frac{c_G^2\tilde\l^3(1-\tilde\l+\tilde\l^2)}{2k_G^2(1-\tilde\l)(1+\tilde\l)^5}
}\ .
\ee
The above is covariant under \eqref{sfjsldjsssk} or equivalently in terms of $k_G$
\be
\label{sfjsldjsssk123}
\boxed{
\tilde\l\to\tilde\l^{-1}\,,\quad k_G\to-k_G}\ .
\ee
We, thus, see that the perturbation theory is naturally organized around the CFT with level $k_G$ deformed by the term  \
$\displaystyle {k_G \tilde \l \ov \pi} J_+ J_-$. 
In fact its covariance is achieved for the two term separately. We expect that this is an exact symmetry to all order in the large $k_G$ expansion.
This can be very useful in trying to extend the $\beta$-function to ${\cal O}(\nicefrac{1}{k_G^3})$ or even to higher ones.

\subsection{Zamolodchikov metric}
\label{group.Zamolo}

Let us consider the two-point correlation function\footnote{
We pass to the Euclidean regime with complex coordinate
$\displaystyle z={1\ov \sqrt{2}} \left(\tau+i\,\sigma\right)$.}
\be
G_{\tilde\lambda}(z_1,\bar z_1;z_2,\bar z_2)=\langle{\cal O}(z_1,\bar z_1){\cal O}(z_2,\bar z_2)\rangle_{\tilde\lambda}\ ,
\ee
where the perturbing current bilinear operator is
\be
\label{sdtadkks}
{\cal O}(z,\bar z)=J^a(z)\bar J^a(\bar z)\, .
\ee
The currents $J^a$ satisfy a current algebra at level $k_G$ with OPEs (operator product expansions)\footnote{Note that we have rescaled the currents as $J^a \to   \nicefrac{J^a}{{\sqrt{k_G}}}$.}
\be
J^a(z_1)J^b(z_2)=\frac{\d_{ab}}{z_{12}^2}+\frac{i}{\sqrt{k_G}}\frac{f_{abc}J^c(z_2)}{z_{12}}\  ,\qquad z_{12}=z_1-z_2\,,
\ee
while the OPE of $J^a$ with $\bar J^a$ is regular.

\no
From \eqref{sdtadkks} we can read off the Zamolodchikov metric as
\be
g(\tilde\l;k)=|z_{12}|^{2(2+\g^{(\cal O)})}G_{\tilde\lambda}(z_1,\bar z_1;z_2,\bar z_2)\,,
\ee
where $\g^{(\cal O)}$ is the anomalous dimension of ${\cal O}$
that is given by \cite{Kutasov:1989dt,Georgiou:2015nka}
\be
\label{anomalouscompositegeneral}
\gamma^{({\cal O})}=2\partial_{\tilde\l}\b^{\tilde\l}(\tilde\l)+\b^{\tilde\l}(\tilde\l)\partial_{\tilde\l}\ln g(\tilde\l;k_G)\,.
\ee

The finite part of the two-point function should behave as
\be
\label{sjfklfhdsjs}
g(\tilde\l;k_G)=\frac12\frac{\text{dim}G}{(1-\tilde\l^2)^2}\left(1+\frac{c_G}{k_G}\frac{Q(\tilde\l)}{(1-\tilde\l)(1+\tilde\l)^3}\right)\,,
\ee
where the zeroth order in the $\nicefrac1k$ expansion was computed in \cite{Kutasov:1989dt,Georgiou:2015nka}.
The poles on the sub-leading part in $\tilde\l=\pm1$ and their order, are chosen such that the line element
\be
\label{lfkfdkdld}
\text{d}\ell^2=g(\tilde\l;k_G)\text{d}\tilde\l^2\,,
\ee
is finite at the PCM and pseudo-dual limits \eqref{limit1} and \eqref{limit2} respectively.
The function $Q(\tilde\l)$ is everywhere analytic with $Q(0)=0$, so that it agrees with the CFT result \cite{Georgiou:2018vbb}
\be
\label{jkshfuswq}
g(0;k_G)=\frac12\text{dim}G\,.
\ee
Demanding that \eqref{lfkfdkdld} is invariant under the symmetry \eqref{sfjsldjsssk123} leads to the condition
\be
\tilde\l^4Q(\tilde\l^{-1})=Q(\tilde\l)\,,
\ee
having as a solution a quartic polynomial of the form
\be
\label{fjjdjkd}
Q(\tilde\l)=\tilde\l\left(c_1+c_2\tilde\l+c_1\tilde\l^2\right)\,,
\ee
where we have used \eqref{jkshfuswq}.
To proceed we note that the Zamolodchikov metric receives no finite
contribution up to  ${\cal O}(\tilde\l^2)$ \cite{Georgiou:2015nka,Georgiou:2016zyo}, fixing $c_{1,2}=0$, Then \eqref{sjfklfhdsjs} simplifies as
\be
\label{sjfklfhdsjsa}
g(\tilde\l;k_G)=\frac12\frac{\text{dim}G}{(1-\tilde\l^2)^2}\, ,
\ee
that is, the possible ${\cal O}(\nicefrac{1}{k_G})$-correction vanishes.

\subsection{$C$-function and the anomalous dimension of the current bilinear}
\label{isoCfunction}

Next we compute the $C$-function from Zamolochikov's $c$-theorem \cite{Zamolodchikov:1986gt} by following 
the procedure introduce in the present context in
\cite{Georgiou:2018vbb}.  We have that \cite{Zamolodchikov:1986gt}
\be
\frac{\text{d}C}{\text{d}t}=\b^i\del_i C=24g_{ij}\b^i\b^j\geqslant0\,.
\ee
For a single coupling $\tilde\l$, the above simplifies to the first order ordinary differential equation
\be
\del_{\tilde\l} C_\text{single}(\tilde\l;k)=24g_{\tilde\l\tilde\l}\b^{\tilde\l}(\tilde\l)\,,\quad g_{\tilde\l\tilde\l}=g(\tilde\l;k_G)\,,
\ee
with solution
\be
\label{djeksiskasjw}
C_\text{single}(\tilde\l;k_G)=c_\text{UV}+24\int_0^{\tilde\l}\text{d}\tilde\l_1\, g(\tilde\l_1;k_G)\b^{\tilde\l}(\tilde\l_1)\,,
\ee
where $c_\text{UV}$ is the central charge at the UV CFT $G_k\times G_k$, namely that
\begin{equation}
c_\text{UV}=2\, \frac{2k\text{dimG}}{2k+c_G}=\text{dimG}\Big(2\, -\frac{c_G}{k_G}\Big)\,.
\end{equation}
Integrating \eqref{djeksiskasjw}, we find that
\be
\label{Cfunctioniso}
C_\text{single}(\tilde\l;k_G)=2\text{dimG}-\frac{c_G\text{dimG}}{k_G}\frac{1+2\tilde\l}{(1-\tilde\l)(1+\tilde\l)^3}-
\frac{3c_G^2\text{dimG}}{2k^2_G}\frac{\tilde\l^4}{(1-\tilde\l)^2(1+\tilde\l)^6}\,.
\ee
This is in agreement with the results of \cite{Georgiou:2018vbb} to leading order in $\nicefrac{1}{k_G}$. In addition, \eqref{Cfunctioniso}
is invariant under \eqref{sfjsldjsssk123} to order $\nicefrac{1}{k_G^2}$, up to a constant
\be
C_\text{single}(\tilde\l^{-1};-k_G)= C_\text{single}(\tilde\l;k_G)+\frac{c_G\,\text{dimG}}{k_G}\,.
\ee
Note the absence of non-analytic terms with branch cuts, i.e. $\ln\frac{1-\l}{1+\l}$, in the expression of the $C$-function. This is due to the choice
of the parameter $d_0=-\nicefrac12$ in
\eqref{fdjfhshdjs} as it has been already noted.
Such terms cannot appear, as it can be seen from a free field expansion around the identity group element \cite{Georgiou:2019aon}.

Finally, we compute the anomalous dimension of ${\cal O}$ to order $\nicefrac{1}{k_G^2}$.
Plugging \eqref{fdjfhshdjs1}, \eqref{sjfklfhdsjsa} into \eqref{anomalouscompositegeneral}, we find that
\be
\label{anomalousisotwo}
\boxed{\gamma^{({\cal O})}=-\frac{2c_G}{k_G}\frac{\tilde\l(1-\tilde\l+\tilde\l^2)}{(1-\tilde\l)(1+\tilde\l)^3}-
\frac{c_G^2}{k_G^2}\frac{\tilde\l^2(3-2 \tilde\l +\tilde\l^2)(1-2 \tilde\l +3\tilde\l^2)}{(1-\tilde\l)^2(1+\tilde\l)^6}
} \ .
\ee
This is in agreement with the results of \cite{Georgiou:2018vbb} to leading order in $\nicefrac1k$ \cite{Georgiou:2015nka}. In addition,
\eqref{anomalousisotwo} is invariant under the symmetry \eqref{sfjsldjsssk123} to order $\nicefrac{1}{k_G^2}$. Again, invariance is achieved for each term separately.

\subsection{Connection with the $\l$-deformed $G_k$}
\label{group.connection}

Let us now consider the $\l$-deformed $\s$-model of $\displaystyle G_k$ 
\cite{Sfetsos:2013wia}.
This model shares the same $\b$-function, Zamolodchikov metric and anomalous dimension as the $\l$-deformed $G_k\times G_k$. The equivalence is based on the perturbation of current algebra
CFTs is driven by the same current bilinears \cite{Georgiou:2017aei}.
However, the UV fixed point differs and its central charge is given by
\begin{equation}
c_\text{UV}=\frac{2k\text{dimG}}{2k+c_G}=\text{dimG}\Big(1-\frac{c_G}{2k_G}\Big)\,.
\end{equation}
Thus, the corresponding $C$-function will be different than \eqref{Cfunctioniso}.
It can be found through \eqref{fdjfhshdjs1}, \eqref{sjfklfhdsjsa} and \eqref{djeksiskasjw}
and reads
\be
\label{Cfunctioniso.original}
C(\tilde\l;k_G)=\text{dimG}-\frac{c_G\text{dimG}}{2k_G}\frac{1+2\tilde\l+2\tilde\l^3+\tilde\l^4}{(1-\tilde\l)(1+\tilde\l)^3}-
\frac{3c_G^2\text{dimG}}{2k^2_G}\frac{\tilde\l^4}{(1-\tilde\l)^2(1+\tilde\l)^6}\,.
\ee
Note that, this is invariant under the symmetry \eqref{cksks}.

\section{The coset space}
\label{coset}

We now turn to the discussion of the coset case.
Let us consider the single level action \eqref{defactigen} for an anisotropic coupling $\l_{ab}$
where now we take the group elements $\frak{g}_{1,2}\in SU(2)$
and $\l_{ab}=\text{diag}(\l,\l,\l_3)$. We would like to compute its RG flow equations at
two-loop order in the $\nicefrac1k$ expansion. It is a tour de force computation, analogue to the one performed in App. \ref{RGequalappend}.
The end result is that the model is renormalizable at order $\nicefrac{1}{k^2}$, there is
no need for a diffeomorphism or an addition of a counter term, and its $\b$-functions read
\ba
\label{betaonetwosu2}
&& \frac{\text{d}\l}{\text{d}t}=-\frac{2\l(\l_3-\l^2)}{k(1+\l_3)(1-\l^2)}-\frac{4\l^3(3\l^2+4\l^4-2\l_3-10\l^2\l_3+5\l_3^2-\l^2\l_3^2+\l_3^4)}{k^2(1+\l_3)^2(1-\l^2)^3}\,,
\nonumber\\
&& \frac{\text{d}\l_3}{\text{d}t}=-\frac{2\l^2(1-\l_3)^2}{k(1-\l^2)^2}+\frac{8\l^2(1-\l_3)^2(\l^4-(3-\l_3)\l_3\l^2+\l_3^2)}{k^2(1+\l_3)(1-\l^2)^4}\,.
\ea
As a consistency check the above result agrees with \eqref{betaonetwo}, in the isotropic limit $\l_3=\l$ and $c_G=4$ for $SU(2)$ in our normalizations.

\no
Let us now consider $\l_3=1$, which is a consistent truncation of the RG flows \eqref{betaonetwosu2}
\be
\label{fjhskdjklswq}
\boxed{
\b^\l(\l)=\frac{\text{d}\l}{\text{d}t}=-\frac{\l}{k}-\frac{4}{k^2}\frac{\l^3}{1-\l^2}
}\ .
\ee
It can be easily seen that \eqref{fjhskdjklswq} is invariant under the symmetry \eqref{cksks} ($c_G=4$)
\begin{equation}
\label{jdkslsnsms}
\l\to\l^{-1}\,,\quad k\to-k-4\,,
\end{equation}
to order $\nicefrac{1}{k^2}.$ This $\b$-function is describing the RG flow between the UV $\l=0$ towards a strongly coupled model at the IR $\l\to1^-$.\footnote{Analyzing the $\b$-function \eqref{fjhskdjklswq} near $\l=1$, we obtain that
\begin{equation*}
\l=1-\frac{\k^2}{k}\,,\quad k\gg1\,,\quad \frac{\text{d}\k^2}{\text{d}t}=1+\frac{2}{\k^2}\,,
\end{equation*}
which matches the two-loop $\b$-function \eqref{fhsjdjss} for the PCM on $S^2$, i.e.
$\text{d}\ell^2=\k^2\left(\text{d}\vartheta^2+\sin^2\vartheta\text{d}\varphi^2\right)$. There is of course an associated limit taken in \eqref{action.coset} which gives the PCM for  $S^2$ and three spectator bosons. This is most easily seem when one goes back to
\eqref{PCMlimit} and sets $E_{33}=0$ since this corresponds to setting $\l_3=1$ as well
as $E_{11}=E_{22}=\k^2$ and $E_{12}=E_{21}=0$. Obviously one may, more generally, have a symmetric space $G/H$ by choosing appropriately
the matrix $E=\diag(\mathbb{I}_{H},\k^2 \mathbb{I}_{G/H})$ in \eqref{PCMlimit}.
}
In what follows, we shall show that $\l_3=1$ corresponds to a parafermionic perturbation of the coset CFT
$\displaystyle \frac{SU(2)_k\times SU(2)_k}{U(1)_k}$, a member of a class of coset CFTs
discussed extensively in \cite{Guadagnini:1987ty}. Let us parametrize the group elements $\frak{g}_{1,2}$ as
\be
\frak{g}_i=\text{e}^{i\s_3\frac{\varphi_i}{2}}\text{e}^{-i\s_2\frac{\vartheta_i}{2}}\text{e}^{i\s_3\frac{\psi_i}{2}}\,,\quad i=1,2\,,
\ee
where $\s_a$ are the Pauli matrices traced normalized to $\text{Tr}(\s_a\s_b)=2\d_{ab}$. Using the above parameterization and Eqs. \eqref{wzwacc} and \eqref{pfkdlddsks}
(with $t_a=\nicefrac{\s_a}{\sqrt{2}}$) into \eqref{defactigen} for $\l_{ab}=\text{diag}(\l,\l,1)$, we find a five dimensional target space $\s$-model since its metric possesses the eigenvector
$
\mathcal{X}=\partial_{\varphi_1}-\partial_{\psi_2}
$, which has vanishing eigenvalue.
To identify the corresponding isometry, we define $\psi=\varphi_1+\psi_2$ and we also relabel $\psi_1\to\varphi_1$,
leading to the $\s$-model
\be
\label{action.coset}
S_\text{coset}=S_\text{CFT}+\frac{k\l}{4\pi}\int\text{d}^2\s\left(\Psi\bar\Psi+\Psi^\dagger\bar\Psi^\dagger\right)\,.
\ee
In the above expression the coset CFT is $\displaystyle\frac{SU(2)_k\times SU(2)_k}{U(1)_k}$, whose metric and the
two-form field read \cite{PandoZayas:2000he}
\be
\text{d}\ell^2=\left(\text{d}\psi+\cos\vartheta_1\text{d}\varphi_1+\cos\vartheta_2\text{d}\varphi_2\right)^2+\text{d}\vartheta_1^2+\sin^2\vartheta_1\text{d}\varphi_1^2
+\text{d}\vartheta_2^2+\sin^2\vartheta_2\text{d}\varphi_2^2
\ee
and
\be
B=\left(\text{d}\psi+\cos\vartheta_1\text{d}\varphi_1\right)\wedge\left(\text{d}\psi
+\cos\vartheta_2\text{d}\varphi_2\right)\, ,
\ee
where we have ignored an overall factor of $\nicefrac{k}{4\pi}$.
The $(\Psi,\bar\Psi)$ are classical expressions for parafermionic operators given by
\be
\label{fhjkj11}
\begin{split}
\Psi=\left(\del_+\vartheta_1+i\sin\vartheta_1\,\del_+\varphi_1\right)\text{e}^{-i(\psi/2+\bar\psi)}\,,\quad
\bar\Psi=\left(\del_-\vartheta_2+i\sin\vartheta_2\,\del_-\varphi_2\right)\text{e}^{-i(\psi/2-\bar\psi)}\,,
\end{split}
\ee
and their complex conjugates $\Psi^\dagger$ and $\bar\Psi^\dagger$ respectively.\footnote{Note that the $\sigma$-model \eqref{action.coset}, 
is invariant under the symmetry:
$\lambda\to-\lambda\,,\quad \psi\to\pi+\psi\,.$
} 
Here $\bar\psi$ represents a non-local function of the angles.
This effectively dresses the operators to ensure conservation $\del_-\Psi=0=\del_+\bar\Psi$.\footnote{In particular, employing the equations of motion \eqref{action.coset} leads for to the non-local function $\bar\psi$ to satisfy
\begin{equation*}
\del_-\bar\psi=\frac12\del_-\psi+\cos\vartheta_2\del_-\varphi_2\ ,\quad \del_+\bar\psi=-\frac12\del_+\psi-\cos\vartheta_1\del_+\varphi_1\,.
\end{equation*}
} 
As a consistency check we have used the action \eqref{action.coset} and the two-loop RG flows  \eqref{trosksk}, \eqref{djrjkdkd}
and derived the $\b$-functions of Eq.\eqref{fjhskdjklswq}. There is no need for
a diffeomorphism or a counter term. Finally, we note the similarity of the \eqref{fhjkj11} to the classical parafermions \cite{Bardacki:1990wj,Bardakci:1990ad} corresponding to the exact coset $SU(2)_k/U(1)_k$ CFT \cite{Fateev:1985mm}.

\subsection{The Zamolodchikov metric}
\label{coset.Zamo}

Similarly to \eqref{sjfklfhdsjs}, the finite part of the two-point function should behave as
\be
\label{sjfklfhdsjsqq}
g(\l;k)=\frac{1}{(1-\l^2)^2}\left(1+\frac{1}{k}\frac{Q(\l)}{1-\l^2}\right)\,,
\ee
where the pole structure in \eqref{sjfklfhdsjsqq}, is inspired from the
$\b$-function in \eqref{fjhskdjklswq}.
Demanding that the line element
\be
\text{d}\ell^2=g(\l;k)\text{d}\l^2\,,
\ee
is invariant under the symmetry \eqref{cksks}, leads to the second degree polynomial
\be
Q(\l)=c_0+c_1\l+c_0\l^2\,.
\ee
The constant $c_0=0$ since the unperturbed Zamolodchikov metric is $k$-independent. The order-$\l$ term also vanishes since it is proportional to correlators involving  and off (three) number of parafermions. Therefore $c_1=0$ as well.
Therefore \eqref{sjfklfhdsjsqq}, is simply given by the $k$-independent part
\be
\label{sjfklfhdsjsqqq}
g(\l;k)=\frac{1}{(1-\l^2)^2}\,.
\ee

\subsection{C-function and the anomalous dimension of the parafermionic bilinear}
\label{cfunction.double.parafermionic}

Similarly to Sec.~\ref{isoCfunction}, $c_\text{UV}$ is the central charge of
the coset CFT $\frac{SU(2)_k\times SU(2)_k}{U(1)_k}$ at $\l=0$, namely
\be
c_\text{UV}=\frac{6k}{k+2}-1=
5-\frac{12}{k}+\frac{24}{k^2}+{\cal O}\left(\frac{1}{k^3}\right)\,,
\ee
and the $C$-function can be found through \eqref{djeksiskasjw}
\be
\label{jdadksdjsks}
C_\text{single}(\l,k)=5-\frac{12}{k}\frac{1}{1-\l^2}+\frac{24}{k^2}\frac{1-2\l^2}{(1-\l^2)^2}\,,
\ee
where we have used \eqref{fjhskdjklswq}, \eqref{sjfklfhdsjsqqq}. It is invariant under the symmetry \eqref{jdkslsnsms} to order $\nicefrac1k^2$, up to an additive constant
\be
C_\text{single}(\l^{-1},-k)= C_\text{single}(\l,k)+\frac{12}{k}-\frac{24}{k^2}\,.
\ee
We are now in position to compute the anomalous dimension of the parafermionic bilinear ${\cal O}$, that was given in \eqref{anomalouscompositegeneral}.
The end result reads
\be
\label{wqoqosksap}
\boxed{
\g^{({\cal O})}=-\frac{2}{k}\frac{1+\l^2}{1-\l^2}-\frac{8}{k^2}\frac{\l^2(3+\l^2)}{(1-\l^2)^2}
}\ .
\ee
which is invariant under the symmetry \eqref{cksks}, to order $\nicefrac1k^2$.
There is a non-trivial check of the above result. Namely that, at the UV CFT point $\l=0$ one
should obtain the exact conformal dimension of the parafermionic bilinear $\D=2+\g^{({\cal O})}=2-\nicefrac2k$, which is indeed the case.

\subsection{Connection with the $\l$-deformed $SU(2)_k/U(1)_k$}
\label{coset.connection}

Let us now consider the $\l$-deformed $\s$-model 
of $\displaystyle SU(2)_k/U(1)_k$ \cite{Sfetsos:2013wia}.\footnote{The two-loop RG equation of this model was also recently considered in \cite{Hoare:2019ark}. The background metric was modified by a quantum correction (determinant) arising from the integration of the gauge fields. 
It was found that the level $k$ runs with the RG scale.} This model shares the same $\b$-function, Zamolodchikov metric and anomalous dimension as
the $\l$-deformed $\displaystyle \frac{SU(2)_k\times SU(2)_k}{U(1)_k}$. The reason is
essentially that the perturbation in both cases is driven by parafermion bilinears which have the same quantum properties, i.e. the same OPE's.
The proof goes along the lines of the similar case in which the perturbation of current algebra
CFTs is driven by the same current bilinears \cite{Georgiou:2017aei}.
However, the UV fixed point differs, so that its central charge is given by
\be
c_\text{UV}=\frac{3k}{k+2}-1=
2-\frac{6}{k}+\frac{12}{k^2}+{\cal O}\left(\frac{1}{k^3}\right)\, .
\ee
Hence, the corresponding $C$-function will be different than \eqref{jdadksdjsks}.
It can be found through \eqref{fjhskdjklswq}, \eqref{sjfklfhdsjsa} and \eqref{djeksiskasjw}
and reads
\be
C(\l,k)=2-\frac{6}{k}\frac{1+\l^2}{1-\l^2}+\frac{12}{k^2}\frac{1-2\l^2-\l^4}{(1-\l^2)^2}\ .
\ee
Note that, this is invariant under the symmetry \eqref{jdkslsnsms}.

\section{Concluding remarks}
\label{conclu}

In this paper we have uncovered an exact symmetry in the space of couplings of the
$\l$-deformed $\s$-models constructed in \cite{Sfetsos:2013wia}.  
This goal was achieved by making use of one of the models constructed in \cite{Georgiou:2016urf,Georgiou:2017jfi}.
More precisely is due to the fact that the single $\l$-deformed model and the doubly $\l$-deformed model with one of the deformation parameters set to zero share the same $\b$-functions to all orders in both the $\l$ and $\nicefrac1k$ expansions \cite{Georgiou:2017aei}. 
For the group case this symmetry is simply stated by \eqref{sfjsldjsssk123}, with the definition \eqref{kgkg}.
Due to its simplicity it is conceivable that we may use it to push the computation of loop-corrections to the $\beta$-function,  operator anomalous dimensions and Zamolodchikov's $C$-function even further. This will be done using also some minimal input form conformal perturbation theory.
This approach seems to be  the most promising way to make progress in this direction since attempting to use the gravitational approach in obtaining loop-corrections higher than two is really cumbersome. Another promising approach could be to use the free field expansion of the  $\l$-deformed action in \cite{Georgiou:2019aon} and study using standard field theoretical methods the renormalization of the interaction vertices. An advantage of this approach is that all the dependence on the deformation parameter $\l$ is already encoded in the vertices. Note that, similar comments hold for the symmetric coset case as well.

\no
We have calculated the anomalous dimensions as exact function of  $\l$ and at two-loops in the $\nicefrac1k$ expansion for the $J\bar J$ composite operator that drives the perturbation
away of the conformal point. We have also calculated Zamolodchikov's $C$-function at the same order. 
 It will be very interesting to extend our results for the single current as well as for
composite current operators of higher rank. In this direction the method developed in \cite{Georgiou:2019jcf} should be useful.

An important comment is in order.
One may wonder if the relation \eqref{kgkg} may get further $\nicefrac1k$-correction with
coefficients that may be $\l$-dependent. Recalling that $k_G$ will be the coefficient in the topological  WZ term, for a well defined theory it has to be an integer. Therefore, since  $k$ is an integer itself such corrections are not expected/allowed.  To conclude, we conjecture that there exists a scheme where the symmetry \eqref{kgkg} persists to all orders in the $\nicefrac{1}{k}$ expansion.

We have also seen that the $\s$-model \eqn{defactigen} is renormalizable without the need to correct the target space geometry, for the case of an isotropic coupling matrix and of an anisotropic coupling for the $SU(2)$ case. For an isotropic coupling matrix this fact was also observed in \cite{Hoare:2019mcc}.

Finally, we quote some a partial result concerning the isotropic deformation of the two-level action \cite{Georgiou:2017jfi}
\be
  S_{k_1,k_2}(\frak{g}_1,\frak{g}_2) = S_{k_1}(\frak{g}_1) + S_{k_2}(\frak{g}_2) + {k\l\ov \pi} \int  \text{d}^2\s\,  {\cal O}\,,\quad {\cal O}=J^a_{1+}\,J^a_{2-}\, ,
\label{defactigendif}
\ee
in which in contrast to \eqref{defactigen} the two levels $k_1$ and $k_2$ are not equal.
In the above action $k=\sqrt{k_1k_2}$ and we also define the parameter $\l_0=\sqrt{\frac{k_1}{k_2}}<1$. 
These models interpolate  between two exact CFTs, namely $G_{k_1}\times G_{k_2}$ at $\l=0$ and $G_{k_2-k_1}\times G_{k_1}$ at $\l=\l_0$ respectively \cite{Georgiou:2017jfi}. The computation performed in App. \ref{RGequalappend} reveals that the model is renormalizable at order $\nicefrac{1}{k^2}$ and there is no need for a diffeomorphism or an addition of a counter term. Its $\b$-functions reads
\be
\label{betaaniso}
\begin{split}
\b^\l(\l;\l_0) & = \frac{\text{d}\l}{\text{d}t}=-\frac{c_G\l^2(\l-\l_0)(\l-\l_0^{-1})}{2k(1-\l^2)^2}\\
&+\frac{c_G^2\l^4(\l-\l_0)(\l-\l_0^{-1})((\l_0+\l_0^{-1})(1+5\l^2)-8\l-4\l^3)}{4k^2(1-\l^2)^5}\,.
\end{split}
\ee
The levels $k_{1,2}$ do not run, thus retaining their topological nature (also) at two-loop order.
For equal levels \eqn{betaaniso} coincides with \eqref{betaonetwo}. Up to ${\cal O}(\nicefrac1k)$  the above expression is invariant under
the symmetry $k_{1,2}\to -k_{2,1}$ and $\l \to \l^{-1}$. Extending this symmetry up to two-loops along the lines
of Sec. \ref{singlegeometry} \, presents some technical challenges and work in direction is in progress.

\section*{Acknowledgements}

We would like to thank Ben Hoare for a useful correspondence.
G. Georgiou and K. Siampos work has received funding from the Hellenic Foundation
for Research and Innovation (HFRI) and the General Secretariat for Research and Technology (GSRT), under grant agreement No 15425.\\
The research of E. Sagkrioti is co-financed by Greece and the European Union (European Social Fund- ESF) through the Operational Programme "Human Resources Development, Education and Lifelong Learning" in the context of the project "Strengthening Human Resources Research Potential via Doctorate Research" (MIS-5000432), implemented by the State Scholarships Foundation (IKY).

\begin{appendices}

\section{Renormalization group flow at two-loops}
\label{RGequalappend}

The scope of this appendix is to work out the RG flow equations of the action
\begin{equation}
S_{k_1,k_2}(\frak{g}_1,\frak{g}_2)=S_{k_1}(\mathfrak{g}_1)+S_{k_2}(\mathfrak{g}_2)+\frac{k\l}{\pi}\int \text{d}^2\s J_{1+}^aJ_{2-}^a\,,\quad k=\sqrt{k_1k_2}\,,
\end{equation}
which is nothing else but the action  \eqref{defactigendif}.
From the above we find the line element
\begin{equation}
\text{d}s^2=R^aR^a+\l_0^{-2}L^{\hat{a}} L^{\hat{a}} +2\l_0^{-1}\l R^aL^{\hat{a}}\,,\quad \l_0=\sqrt{\frac{k_1}{k_2}}\,,
\end{equation}
and the two-form
\begin{equation}
B=B_0+\l_0^{-1}\l R^a\wedge L^{\hat a}\ ,
\end{equation}
where $B_0$ is the two-form which corresponds to the two WZW models at levels $k_{1,2}$ with
\be
H_0=\text{d}B_0=-\frac16 f_{abc}\left(R^a\wedge R^b\wedge R^c+\l_0^{-2} L^{\hat a}\wedge L^{\hat b}\wedge L^{\hat c}\right)\,.
\ee
In the above we have disregarded an overall $\frac{k_1}{2\pi}$ factor and the Maurer--Cartan one forms are given by
\begin{equation}
\begin{split}
R^a=-i\text{Tr}(t^a\text{d}\frak{g}_1\frak{g}_1^{-1}),\qq L^{\hat{a}}=-i\text{Tr}(t^a\frak{g}^{-1}_2\text{d}\frak{g}_2)\,,\\
\text{d}R^a=-\frac{1}{2}f_{abc}R^{{b}}\wedge R^{{c}},\qq \text{d}L^{\hat{a}}=\frac{1}{2}f_{abc}L^{\hat{b}}\wedge L^{\hat{c}}\,.
\end{split}
\end{equation}
Here, the unhatted and hatted indices denote the Maurer--Cartan one forms evaluated at the group elements $\frak{g}_1$ and $\frak{g}_2$ respectively. By introducing the vielbeins
\begin{equation}
\text{e}^a=R^a,\quad \text{e}^{\hat{a}}=\l R^a+\l_0^{-1}L^{\hat{a}} 
\label{veilbeins}
\end{equation}
and the double index notation $A=(a,\hat{a})$, the line element can be written as
\begin{equation}
\label{fskskfhjsk}
\text{d}s^2=(1-\l^2) \text{e}^a\text{e}^a+\text{e}^{\hat{a}}\text{e}^{\hat{a}}=G_{AB}\,\text{e}^A\text{e}^B\,.
\end{equation}
The spin connection and the torsion for the action  \eqref{defactigendif} have been found in Eqs. (2.14) and (2.16) of \cite{Sagkrioti:2018rwg}. For an isotropic coupling $\l_{ab}=\l\d_{ab}$ read
\begin{align}
\begin{split}
&\omega_{ab}=-\frac{1}{2}(1-\l^2)f_{abc}\text{e}^c+\frac{\l}{2}(1-\l_0\l)f_{abc}\text{e}^{\hat{c}}\,,\\
&\omega_{\hat{a}b}=\omega_{a\hat{b}}=\frac{\l}{2}(\l_0\l-1)f_{abc}\text{e}^c\,,\\
&\omega_{\hat{a}\hat{b}}=-\l_0\l f_{abc}\text{e}^c+\frac{\l_0}{2}f_{abc}\text{e}^{\hat{c}}\ ,
\end{split}
\end{align}
where we note that, since the metric \eqref{fskskfhjsk} is constant, $\omega_{AB}$ is antisymmetric.
Also
\begin{align}
\begin{split}
&H=-\frac{1}{6}\left(1-\l^2(3-2\l_0\l)\right)f_{abc}\,\text{e}^a\wedge{}\text{e}^b\wedge \text{e}^c\\
&\phantom{xxx}-\frac{\l}{2}(1-\l_0\l)f_{abc}\,\text{e}^{\hat{a}}\wedge \text{e}^b \wedge \text{e}^c-\frac{\l_0}{6}f_{abc}\,\text{e}^{\hat{a}}\wedge \text{e}^{\hat{b}}\wedge \text{e}^{\hat{c}}\,.
\end{split}
\end{align}
For the two-loop computation, we are going to need the torsionfull spin connection $\om^-_{AB}$
\be
\omega^-_{AB}=\omega^-_{AB|C}\text{e}^C=\left(\omega_{AB|C}-\frac12\,H_{ABC}\right)\text{e}^C
\ee
and in terms of components is given by \cite{Sagkrioti:2018rwg}
\begin{align}
\begin{split}
&\omega^-_{ab}=\l^2(\l_0\l-1)f_{abc}\text{e}^c+\l(1-\l_0\l)f_{abc}\text{e}^{\hat{c}}\,,\\
&\omega^-_{\hat{a}b}=\omega^-_{a\hat{b}}=0\,,\\
&\omega^-_{\hat{a}\hat{b}}=-\l_0\l f_{abc}\text{e}^c+\l_0f_{abc}\text{e}^{\hat{c}}\,.
\end{split}
\end{align}
We can now compute the torsionfull Riemann two-form $\Om^-_{AB}$ reads
\begin{equation}
\Om^-_{AB}=\frac{1}{2}R^-_{ABCD}\,\text{e}^C\wedge \text{e}^D=\text{d}\omega^-_{AB}+\omega^-_{AC}\wedge \omega^{-C}{}_B\ 
\end{equation}
and the corresponding components read
\begin{equation}
R^-_{ABCD}=\left(\omega^K{}_{C|D}-\omega^K{}_{D|C}\right)\omega^-_{AB|K}+\omega^-_{AK|C}\omega^{-K}{}_{B|D}-
\omega^-_{AK|D}\omega^{- K}{}_{B|C}\,, \label{Riemann}
\end{equation}
where we have used that $\omega_{AB|C}$'s are constants.
Employing the above and \eqref{Riemann}, we find the components of the torsionfull Riemann tensor
\be
\begin{split}
&R^-_{abcd}=R_1 f_{abe}f_{cde}\ ,
\quad R^-_{abc\hat{d}}=R_2 f_{abe}f_{cde}\ ,\quad R^-_{ab\hat{c}\hat{d}}=R_3 f_{abe}f_{cde}\ ,
\\
&R_1=\l^3\L\,,\quad R_2=-\l^2\L\ ,
\quad R_3=\l\L\ ,
\quad \L=\frac{(\l-\l_0)(\l_0\l-1)}{1-\l^2}\ .
\end{split}
\ee
While the other components identically vanish.
We are also going to need $H^2_{AB}=H_{ACD}H_B{}^{CD}$, where
\begin{align}
\begin{split}
&(H^2)_{ab}=c_GH_1\d_{ab}\ ,\qq H_1=\frac{1-4\l^2+\l^4\left(7+2\l_0\big(\l_0-\l(4-\l_0\l)\big)\right)}{(1-\l^2)^2}\ ,
\\
&(H^2)_{\hat{a}b}=c_GH_2\d_{ab}\ , \qq H_2=\frac{\l(1-\l_0\l)\big(1-\l^2(3-2\l_0\l)\big)}{(1-\l^2)^2}\ ,
\\
&(H^2)_{\hat{a}\hat{b}}=c_GH_3\d_{ab} \ , \qq H_3=\frac{\l^2(1-\l_0\l)^2+\l_0^2(1-\l^2)^2}{(1-\l^2)^2}\ .
\end{split}
\end{align}
We are now in position to compute the two loop $\b$-functions of \eqref{defactigendif}.
These were given by
\be
\label{trosksk}
\frac{\text{d}}{\text{d}t}\left(G_{MN}+B_{MN}\right)=\left(\b^{(1)}_{AB}+\b^{(2)}_{AB}\right)\text{e}^A{}_M\text{e}^B{}_N\,,
\ee
where $t=\ln\mu^2$, $\mu$ is the RG scale and
\cite{Curtright:1984dz,Braaten:1985is,Hull:1987pc,Hull:1987yi,Metsaev:1987bc,Metsaev:1987zx,Osborn:1989bu}\footnote{We are using Eq.(7) in Hull--Townsend \cite{Hull:1987pc}  or equivalently Eq.(4.26) in Osborn \cite{Osborn:1989bu}. Note that in our conventions of the generalized Riemann tensor we replace $+\to -$ and we also rescale $H \to \nicefrac12 H$, due to our different normalization of the $H = \text{d}B$ field.}
\be
\label{djrjkdkd}
\b^{(1)}_{AB}=R^-_{AB}\,,\quad
\b^{(2)}_{AB}=R^-_{ACDE}\left(R^{-CDE}{}_B-\frac{1}{2}R^{-DEC}{}_B\right)+\frac{1}{2}(H^2)^{CD}R^-_{CABD}\,.
\ee

To proceed we analyze the left-hand side of \eqref{trosksk}, which equals to
\be
\label{qiakmd}
\frac{\text{d}}{\text{d}t}\left(G_{MN}+B_{MN}\right)=2\frac{\text{d}\l}{\text{d}t}\left(\text{e}^a{}_M\text{e}^{\hat{a}}{}_N-\l\,\text{e}^a{}_M\text{e}^a{}_N\right)\ .
\ee
The one-loop contribution $\b^{(1)}$ was analyzed in \cite{Sagkrioti:2018rwg} and we shall present the end result
\be
\label{qiakmd1}
\b^{(1)}_{ab}=c_G\d_{ab}\frac{R_1}{1-\l^2}\,,\quad
\b^{(1)}_{\hat{a}b}=\b^{(1)}_{\hat{a}\hat{b}}=0\,,\quad
\b^{(1)}_{a\hat{b}}=c_G\d_{ab}\frac{R_2}{1-\l^2}\,,
\ee
with $\b^{(1)}_{ab}=-\l\b^{(1)}_{a\hat{b}}$.	
Then, we move to the two-loop contribution $\b_{AB}^{(2)}$. Employing the above results we find\footnote{Where we have used the identity
$f_{aa_1a_2}f_{ba_2a_3}f_{ca_3a_1}=\frac{c_G}{2}f_{abc}$, easily proved using the Jacobi identity.}
\begin{align}
\label{qiakmd2}
\begin{split}
&\b^{(2)}_{ab}=c_G^2\d_{ab}\left(\frac{R_1^2}{(1-\l^2)^3}+\frac{1}{2}\frac{R_2^2-H_1R_1}{(1-\l^2)^2}-\frac{1}{2}\frac{H_2R_2}{1-\l^2}\right)\ ,
\\
&\b^{(2)}_{\hat{a}b}=\b^{(2)}_{\hat{a}\hat{b}}=0\ ,
\\
&\b^{(2)}_{a\hat{b}}=c_G^2\d_{ab}\left(\frac{R_1R_2}{(1-\l^2)^3}+\frac{1}{2}\frac{R_2R_3-H_1R_2}{(1-\l^2)^2}-\frac{1}{2}\frac{H_2R_3}{1-\l^2}\right)\ ,
\end{split}
\end{align}
where $\b^{(2)}_{ab}=-\l\b^{(2)}_{a\hat{b}}$.
Employing \eqref{qiakmd}, \eqref{qiakmd1}, \eqref{qiakmd2} into \eqref{trosksk} and reinserting the overall $k_1$
factors on the line element and two-form field, one finds
\begin{align}
\label{betaonetwoappendaniso}
\begin{split}
\b^\l(\l;\l_0)=& \frac{\text{d}\l}{\text{d}t}=-\frac{c_G}{2k}\frac{\l^2(\l-\l_0)(\l-\l_0^{-1})}{(1-\l^2)^2}\\
&+\frac{c_G^2}{4k^2}\frac{\l^4(\l-\l_0)(\l-\l_0^{-1})\big((\l_0+\l_0^{-1})(1+5\l^2)-8\l-4\l^3\big)}{(1-\l^2)^5}\ 
\end{split}
\end{align}
and the levels $k_{1,2}$ do not flow.

\subsection{Equal levels}
\label{limits}

For equal levels $k_1=k=k_2$, \eqref{betaonetwoappendaniso} drastically simplifies to
\be
\label{betaonetwoappend}
\b^\l(\l)={\text{d}\l\ov \text{d}t}=  -{c_G\ov 2 k} {\l^2\ov (1+\l)^2} +
{c_G^2\ov 2 k^2 } {\l^4(1-2\l)\ov (1-\l)(1+\l)^5} \ .
\ee
Let us now analyze two interesting limits of the above expression around $\l=1$ and $\l=-1$ for $k\to\infty$ -- retaining its topological
nature at two-loop in $\nicefrac{1}{k}$ expansion.
These limits were studied in detail in Sec.~\ref{singlegeometry} and they correspond to the isotropic PCM and the pseudo-dual chiral model respectively.\footnote{Analogue limits exist for the single $\l$-deformed model  \cite{Sfetsos:2013wia,Georgiou:2016iom},
corresponding to the non-abelian T-dual of the isotropic PCM and the pseudo-dual chiral model respectively.}
In particular, expanding around $\lambda=1$ and $\lambda=-1$ one finds
\begin{equation}
\label{limit1}
\lambda=1-\frac{\kappa^2}{k}\ ,\quad k\gg1\,,\quad {\text{d}\kappa^2\ov \text{d}t}=\frac{c_G}{8}+\frac{c_G^2}{64\kappa^2}\ 
\end{equation}
and
\begin{equation}
\label{limit2}
\lambda=-1+\frac{1}{b^{2/3}k^{1/3}}\ ,
\quad k\gg1\ ,
\quad {\text{d}b\ov \text{d}t}=\frac{3}{4}\,c_Gb^3-\frac{9}{8}c_G^2b^5\ .
\end{equation}
In what follows, we shall prove that the above limiting expressions are in agreement with those found from the PCM and the pseudo-dual chiral model:
Let us consider the action \eqref{PCMlimit} for an isotropic PCM with $E_{ab}=2\k^2\d_{ab}$, where $\k$ is a coupling constant.
This is a pure metric non-linear $\sigma$-model, whose $\b$-functions drastically simplify to \cite{Ecker:1972bm,Honerkamp:1971sh,Friedan:1980jf,Friedan:1980jm}:
\begin{equation}
\label{fhsjdjss}
\frac{\text{d}G_{\mu\nu}}{\text{d}t}=R_{\mu\nu}-R_{\mu\kappa\rho\sigma}R^{\rho\sigma\kappa}{}_\nu\ ,
\end{equation}
where $G_{\mu\nu}=2\kappa^2R^a_\mu R^a_\nu$. Using of the above we easily find
\begin{equation}
{\text{d}\kappa^2\ov \text{d}t}=\frac{c_G}{8}+\frac{c_G^2}{64\kappa^2}\ ,
\end{equation}
which is in agreement with \eqref{limit1}.

\no
Let us now consider the action \eqref{pseudolimit} for the pseudo-dual chiral model \cite{Nappi:1979ig},
with $\displaystyle G_{ab}=\frac{\delta_{ab}}{2b^{2/3}}$ and
$\displaystyle B_{ab}=\frac{1}{6}f_{abc}v^c$. This is a torsionfull $\s$-model
whose $\b$-functions were given in \eqref{trosksk}, \eqref{djrjkdkd}.
Using the above, one finds
\be
 {\text{d}b\ov \text{d}t}=\frac{3}{4}\,c_Gb^3-\frac{9}{8}c_G^2b^5\,,
\ee
which is in agreement with \eqref{limit2}.

\end{appendices}

\end{document}